\documentclass{iau}

\pdfoutput=1

\usepackage{graphicx}
\DeclareGraphicsExtensions{.pdf}
\usepackage[OT4]{fontenc}

\usepackage{url}

\title[Mapping the Cosmic Web with the largest all-sky surveys]
{Mapping the Cosmic Web\\with the largest all-sky surveys}

\author[Bilicki, Peacock, Jarrett, Cluver \& Steward]
{Maciej Bilicki$^{1,2}$,
John A. Peacock$^3$,
Thomas H. Jarrett$^1$,\\
Michelle E. Cluver$^1$
\and Louise Steward$^1$}

\affiliation{$^1$Department of Astronomy, University of Cape Town, South Africa\\
email: {\tt maciek(at)ast.uct.ac.za}\\[\affilskip]
$^2$Kepler Institute of Astronomy, University of Zielona G\'{o}ra, Poland\\[\affilskip]
$^3$Institute for Astronomy, University of Edinburgh, United Kingdom}

\pubyear{2014}
\volume{308}
\jname{The Zeldovich Universe: Genesis and Growth of the Cosmic Web}
\editors{R. van de Weygaert, S. Shandarin, E. Saar \& J. Einasto, eds.}

\begin{document}
\maketitle

\begin{abstract}
Our view of the low-redshift Cosmic Web has been revolutionized by galaxy redshift surveys such as 6dFGS, SDSS and 2MRS. However, the trade-off between depth and angular coverage limits a systematic three-dimensional account of the entire sky beyond the Local Volume ($z<0.05$). In order to reliably map the Universe to cosmologically significant depths over the full celestial sphere, one must draw on multiwavelength datasets and state-of-the-art photometric redshift techniques. We have undertaken a dedicated program of cross-matching the largest photometric all-sky surveys -- 2MASS, WISE and SuperCOSMOS -- to obtain accurate redshift estimates of millions of galaxies. The first outcome of these efforts -- the 2MASS Photometric Redshift catalog (2MPZ, \cite[Bilicki \etal\ 2014a]{2MPZ}) -- has been publicly released and includes almost 1~million galaxies with a mean redshift of $z=0.08$. Here we summarize how this catalog was constructed and how using the WISE mid-infrared sample together with SuperCOSMOS optical data allows us to push to redshift shells of $z\sim0.2$--$0.3$ on unprecedented angular scales. Our catalogs, with $\sim20$ million sources in total, provide access to cosmological volumes crucial for studies of local galaxy flows (clustering dipole, bulk flow) and cross-correlations with the cosmic microwave background such as the integrated Sachs-Wolfe effect or lensing studies.
\end{abstract}

\firstsection 
\section{The need for all-sky galaxy surveys in three dimensions}
A complete picture of the Cosmic Web we live in would only be known if we could observe the entire extragalactic sky ($4\pi$ sr) up to the surface of last scattering and with full redshift information. This is of course observationally unachievable, in part due to our Galaxy creating the so-called Zone of Avoidance (ZoA) and the instrumental limitations which force us to choose between wide angular coverage and the depth of a redshift survey. There is, however, a reason and need to observe as much of the sky as possible in three dimensions, as only this enables the framework for comprehensively testing cosmological models. For instance, cosmic microwave background (CMB) observations provide a consistent picture of an early Universe which was very homogeneous and isotropic; such an assumption is also applied to the late-time Universe when modeling it with a simple Friedman-Lema\^{\i}tre metric. However, validity of this Copernican Principle (CP) needs to be verified observationally, and this is only possible if we observe the deep cosmos in all possible directions, and at various redshifts: cosmological tests of the CP require access to the entire celestial sphere in 3D. For instance, analyses of the CMB data have brought to light several `anomalies' such as the quadrupole-octopole alignment or low observed variance of the CMB signal (\cite[Planck collaboration XXIII 2013]{PlanckXXIII}). A fundamental question is whether these anomalies are confirmed as today's anisotropy and/or inhomogeneity in the galaxy distribution; some studies indicate this is not the case (\cite[Hirata 2009; Pullen \& Hirata 2010]{Hirata09,PH10}). The issue, however, requires more scrutiny using more comprehensive low-redshift catalogs. One possible manifestation of late-time violation of the CP would be high-amplitude large-scale flows of galaxies and indeed claims of their existence have been made (\cite[Kashlinsky \etal\ 2008; Watkins \etal\ 2009]{Kashlinsky,WFH09}), although several independent analyses show otherwise (e.g.\ \cite[Nusser \& Davis 2011; Turnbull \etal\ 2012]{ND11,Turnbull}). Also the issue of which structures, and on what scales, contribute to the pull on the Local Group of galaxies has still not been settled; despite some claims of the convergence of the `clustering dipole' at distances $\sim100$ Mpc (\cite[Rowan-Robinson \etal\ 2000; Erdogdu \etal\ 2006]{RR00,Erdogdu06}), other analyses suggest influence on our motion from much larger depths (\cite[Kocevski \& Ebeling 2006; Bilicki \etal\ 2011]{KE06,BCJM11}; see also \cite[Nusser \etal\ 2014]{NDB14}). Finally, several other important cosmological signals -- such as the integrated Sachs-Wolfe effect (ISW), CMB lensing on the large-scale structure (LSS) or baryon acoustic oscillations (BAO) -- do not require full-sky coverage, but are only detectable with sufficiently wide-angle galaxy surveys probing large volumes.
\vspace{1mm}

\textbf{\textit{State of the art in all-sky surveys of galaxies.}}
Presently, the largest all-sky catalog of extended sources is the Two Micron All Sky Survey Extended Source Catalog (2MASS XSC, \cite[Jarrett \etal\ 2000]{2MASS.XSC}). It contains approx.\ 1.5 million galaxies, of which 1~million are within the completeness limit of the survey, $K_s<13.9$ mag. This catalog is however only photometric, and the largest all-sky 3D spectroscopic dataset is its subsample, the 2MASS Redshift Survey of 44,000 galaxies (2MRS, \cite[Huchra \etal\ 2012]{2MRS}), with median redshift of only $\langle z \rangle=0.03$. There have been attempts to expand 2MRS by adding redshifts of 2MASS galaxies available from other surveys such as 6dFGS or SDSS, which resulted in the the 2M++ compilation by \cite{LH11} of 70,000 galaxies; the depth of such a sample is however non-uniform and limited to what is available from 2MRS on the parts of the sky not covered by other redshift surveys.

In addition to 2MASS, there exist two other major photometric all-sky datasets, namely the Wide-field Infrared Survey Explorer (WISE, \cite[Wright \etal\ 2010]{WISE}) and optical SuperCOSMOS (SCOS for short, \cite[	Hambly \etal\ 2001]{SCOS}). These samples include hundreds of millions sources, mostly galaxies at high latitudes. WISE mid-infrared data were gathered by an orbiting telescope, which makes them free of such issues as seeing or atmospheric glow. WISE currently provides only a point-source catalog with no `XSC' release; however, its limited angular resolution will not allow to resolve sources beyond the depths already probed by 2MASS (\cite[Cluver \etal\ 2014]{Cluver14}). SCOS on the other hand was produced from digitized and calibrated photographic plate data, which are of limited accuracy compared to modern CCD-based surveys. Despite such shortcomings, these datasets will remain the largest all-sky astronomical catalogs at least until Gaia, which will however be dominated by stars. We have thus decided to take advantage of the information contained in 2MASS, WISE and SCOS, focusing on their usefulness for extragalactic science. Specifically, we have undertaken a dedicated program of combining them into multi-wavalength galaxy datasets and adding the third dimension through the methodology of photometric redshifts.

\section{2MASS Photometric Redshift catalog (2MPZ)}
\begin{figure}[!t]
\centering
\includegraphics[width=\textwidth,height=69mm]{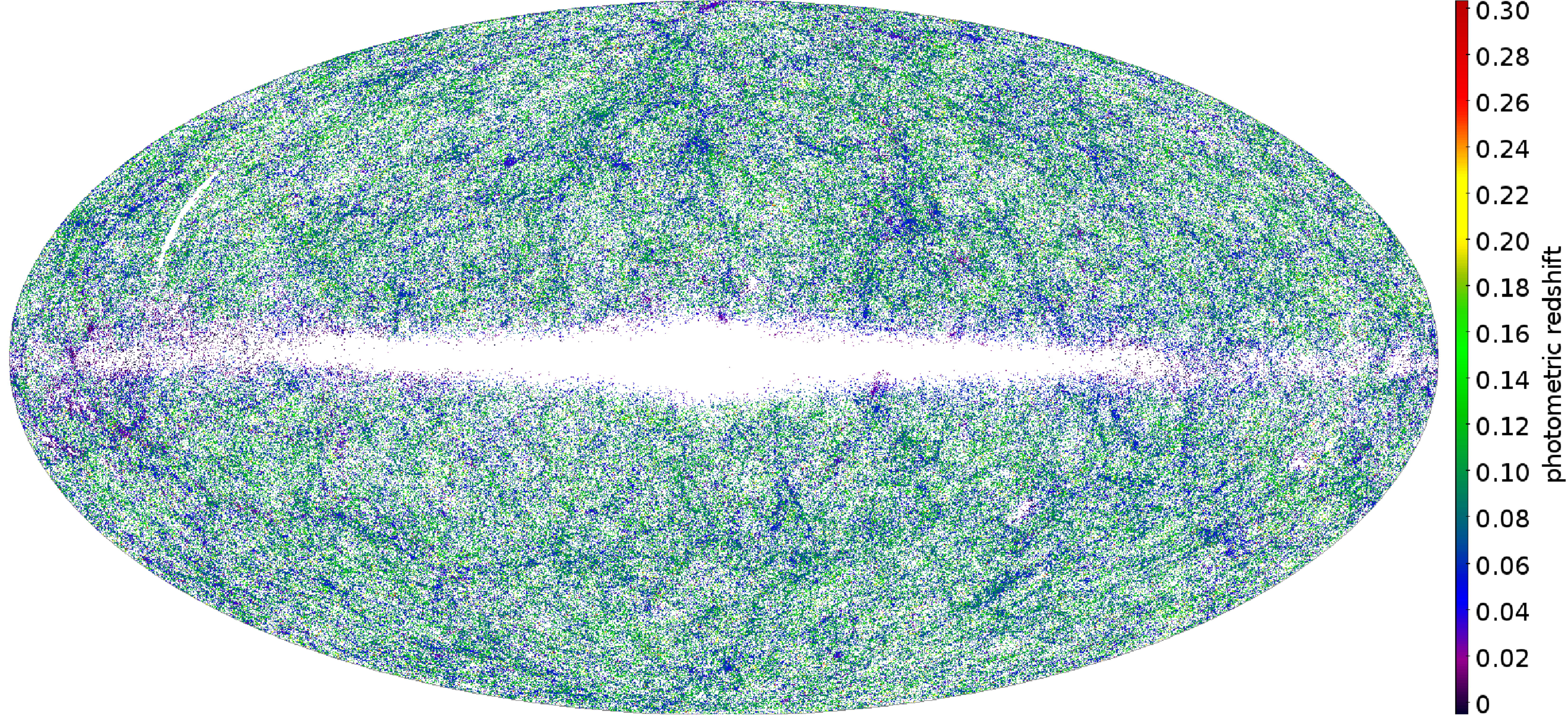}
\caption{All-sky Aitoff projection of 1 million galaxies in the 2MASS Photometric Redshift catalog (2MPZ) in Galactic coordinates, color-coded by photometric redshifts.}
\label{Fig: 2MPZ Aitoff}
\end{figure}

At the moment, about 30\% of 2MASS galaxies have spectroscopic redshifts from other surveys, such as 2MRS, 2dFGRS, 6dFGS or SDSS. This percentage is likely to improve in the coming years, notably in the southern hemisphere with the planned TAIPAN survey, we cannot, however, hope for \textit{complete} spectroscopic coverage of all the 2MASS galaxies in the coming years. Presently, the only way to obtain 3D information for this sample -- even if of limited accuracy -- is to estimate redshifts based on other information, such as fluxes,. This is the widely used technique of \textit{photometric redshifts} (photo-z's), which so far has not been very popular for low-$z$ samples because of limited availability of multiwavelength data. The situation has though changed mostly due to WISE, which together with SCOS allowed us to add the third dimension to the 2MASS XSC sample. For that purpose we used 2MASS $J$, $H$ and $K_s$ near-IR bands, together with mid-IR WISE ($W1$, 3.4 $\mu$m and $W2$, 4.6 $\mu$m) and optical $B, R, I$ from SCOS. This gave us 8-band coverage for almost 95\% of 2MASS galaxies (with incompleteness mostly at low Galactic latitudes), and the availability of comprehensive spectroscopic subsamples allowed to apply empirical (machine-learning) methodology to calculate photo-z's. We employed an artificial neural network algorithm (ANNz, \cite[Collister \& Lahav 2004]{ANNz}), trained on a representative spec-z subsample of 350,000 galaxies. As a result, we have produced the 2MASS Photometric Redshift catalog (2MPZ, \cite[Bilicki \etal\ 2014a]{2MPZ}), containing 940,000 galaxies with $\langle z \rangle = 0.07$, covering most of the sky. This dataset was publicly released in December 2013 and is available for download from  \url{http://surveys.roe.ac.uk/ssa/TWOMPZ}. The photometric redshifts are well constrained: $1\sigma$ scatter in $\delta z$ is $\sigma_{\delta z}=0.015$, typical photo-z error is 13\% and there are only 3\% outliers beyond $3\sigma_{\delta z}$. Fig.\ \ref{Fig: 2MPZ Aitoff} presents the all-sky distribution of 2MPZ sources, color-coded by photometric redshifts. Despite the tendency of the photo-z's to dilute radial information, the 3D cosmic web is evident, with major structures such as the Shapley Superconcentration prominent in the maps. Note that by using infrared-selected galaxies we are able to probe deep into the ZoA, although photo-z's at very low latitudes are of limited accuracy due to uncertainties in extinction and large star density compromising the photometric multiwavelength data.

\vspace{1mm}
\textbf{\textit{First cosmological results from 2MPZ.}}
The 2MPZ catalog has been already applied to several cosmological tests, of which some have been published, and others are in preparation. \cite{AS14} used it to test the degree of isotropy in the Local Universe through the $K_s$-band luminosity function (LF), while \cite{XWH14} showed that the sample is appropriate for galaxy cluster identification. In a forthcoming paper (\cite[Steward \etal\ 2014, in prep.]{Steward14}) we will present the constraints on the ISW obtained by cross-correlating 2MPZ with Planck CMB data. Working in redshift shells (cf.\ \cite[Francis \& Peacock 2010, FP10]{FP10}), we have found only mild preference for the ISW over no signal: less than $2\sigma$ (odds 3.5:1). This is an improvement over FP10 (who had odds 1.5:1), possible thanks to our much more accurate photo-z's, the lack of significant detection is however hardly surprising. 2MASS is too shallow to probe the full ISW signal, and the constraints would not be much more significant even if we had full spectroscopic coverage.

Other applications in progress of the 2MPZ data include a detailed study of the near-IR LF at $z\sim0$ (cf.\ \cite[Branchini \etal\ 2012]{BND12}) from the largest sample to date (Feix \etal, in prep.), as well as constraining the bulk flow in a sphere of $\sim300$ Mpc/$h$ from LF variations (cf.\ \cite[Nusser \etal\ 2011; Feix \etal\ 2014]{NBD11,FND14} and Feix \etal\ in this volume). The catalog is now also analyzed in terms of its applicability to various clustering-related analyses: looking for the transition to homogeneity through the angular correlation function (cf. \cite[Alonso \etal\ 2014]{Alonso14}) or more general attempts to recover the 3D power spectrum and correlation function from photometric redshifts, setting the ground for upcoming and future surveys such as DES or Euclid. Many more applications of the catalog are of course expected.

\section{Beyond 2MASS: 20 million galaxies from WISE$\times$SuperCOSMOS}

The depth of 2MPZ is limited by the shallowest of the three cross-matched samples, i.e.\ 2MASS. Both SCOS and WISE probe the Cosmic Web to much larger distances: the former is about 3 times deeper than 2MASS, while the latter reaches even farther. In addition, both contain a significant number of quasars, some at high redshifts (especially WISE). By pairing up WISE with SCOS we were thus able to access sources and redshift shells not available with 2MPZ. Here the sample is limited by the depth and accuracy of the optical data: we adopted $B<21$ and $R<19.5$ (AB) as the reliability limits, requiring also source detections in both these bands, while WISE was preselected to have measurements in $W1$ and $W2$. Cross-matching the two samples at $|b_\mathrm{Gal}|>10^\circ$ gives 170 million sources, however such a catalog is dominated by stars at the bright end and at low latitudes, so further cuts were needed to clean up the sample. Using Galaxy  And Mass Assembly (GAMA, \cite[Driver \etal\ 2009]{GAMA}) and SDSS DR10 (\cite[Ahn \etal\ 2014]{SDSS.DR10}) spectroscopic data, we defined simple color cuts to purify the sample of stars and high-$z$ quasars, which together with SCOS morphological information gave us about 20 million galaxies over $>3\pi$ sr. The quasars present in our sample, which were treated as contamination here, will be a subject of a future study; we estimate that there might be even 2 million of them in the WISE$\times$SCOS cross-match.  A forthcoming paper (Krakowski \etal\ 2014) will present a more sophisticated method of source classification in our catalog, namely with the use of Support Vector Machines (e.g. \cite[Solarz \etal\ 2012; Ma{\l}ek \etal\ 2013]{Solarz12,Malek13}) trained on SDSS DR10 spectroscopic data.

\begin{figure}[!t]
\centering
\includegraphics[width=1.04\textwidth,height=74mm]{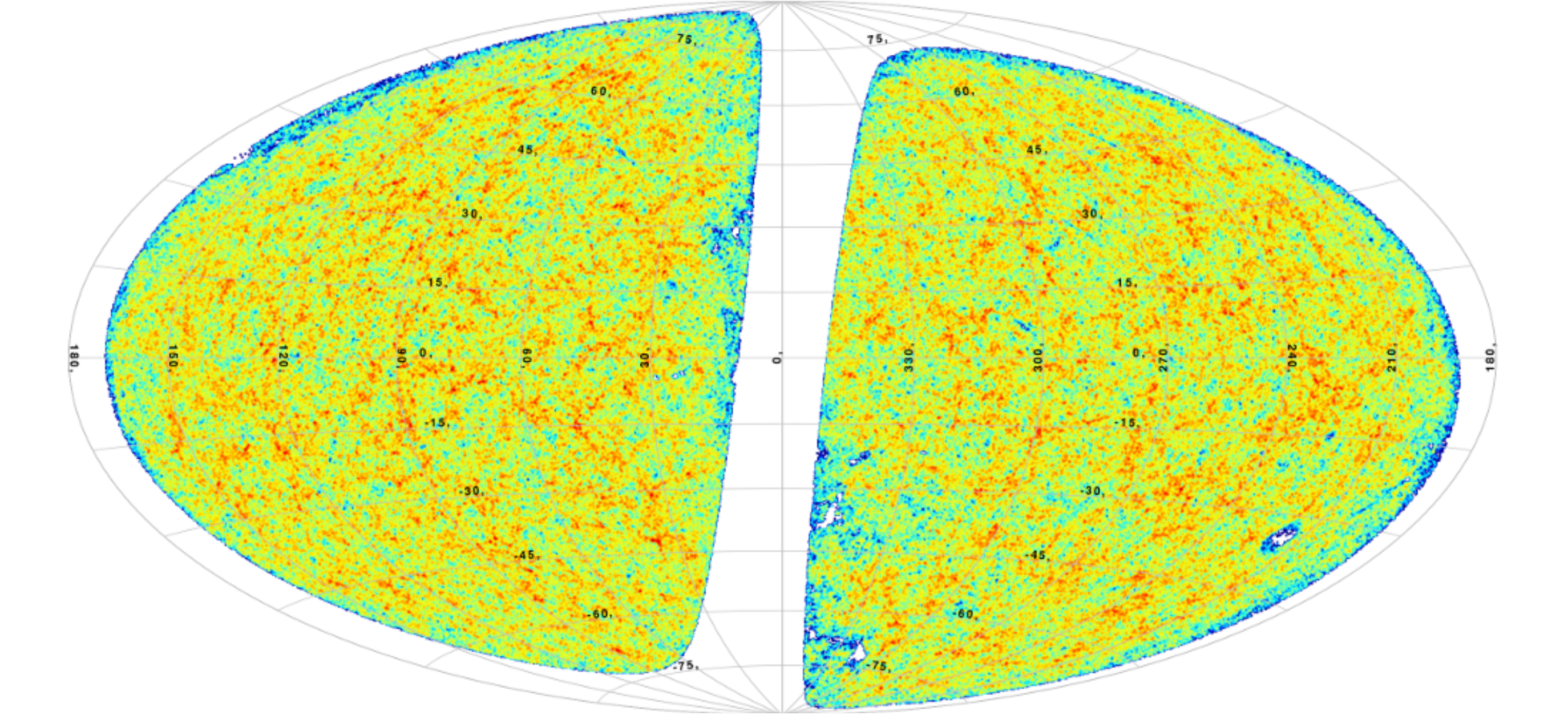} 
\vspace*{0.5mm}
\caption{The Cosmic Web 2.5 Gyr ago: the large-scale structure of the Universe at $z=0.2$. The plot shows 1.7 million galaxies from the WISE$\times$SCOS photometric redshift catalog in a shell of $0.19 < z_\mathrm{phot} < 0.21$, in Aitoff projection in Supergalactic coordinates. The gaps are due to our Galaxy and the LMC obscuring the view.}
\label{Fig: LSS at z=0.2}
\end{figure}

The next step was to calculate photometric redshifts for the galaxies in the sample. We used five photometric bands for that purpose: optical $B,R$ and infrared $W1,W2,W3$. As in 2MPZ, we took the empirical approach (ANNz, \cite[Collister \& Lahav 2004]{ANNz}) but here the training sets were the most recent GAMA and SDSS DR10 data. The former of the two spectroscopic catalogs is crucial here, as it is complete in 3 equatorial fields to $r<19.8$, which makes it deeper than our sample, hence representative for photo-z calibration. The resulting photometric redshifts have a median $\langle z \rangle \sim 0.2$, the redshift distribution is however broad and the sample probes the LSS reliably up to $z\sim 0.35$, with some information even at $z\gtrsim0.4$. Normalized $1\sigma$ scatter of photo-z errors is $\sigma_{\delta z}=0.03$, median error is 12\% and there are 3\% outliers over $3\sigma_{\delta z}$. With the present catalog this is the best performance attainable, as there is no other all-sky data of sufficient depth that could bring additional bands to our full sample. Further details regarding the WISE$\times$SCOS photometric redshift catalog construction will be provided in a forthcoming paper (\cite[Bilicki \etal\ 2014b, in prep.]{WISCAZ}). Fig.\ \ref{Fig: LSS at z=0.2} shows an example of what the WISE$\times$SCOS photo-z sample gives access to: the Cosmic Web as it was 2.5 Gyr ago. Pictured here are 1.7 million galaxies in a redshift shell of $0.19 < z_\mathrm{phot} < 0.21$; note that there are more sources in this narrow shell than in the entire 2MASS galaxy catalog. 

\vspace{1mm}
\textbf{\textit{Possible applications of the WISE$\times$SuperCOSMOS photo-z catalog.}} The availability of the deep, almost all-sky WISE$\times$SCOS data opens new possibilities of LSS analysis, as this is the first 3D galaxy catalog which probes the redshifts of $z=0.2\sim0.3$ on 75\% of the sky. In particular, it will be suitable for similar studies as done with 2MPZ, but in more distant redshift shells and in much larger volumes. Most generally, we expect it to allow for testing the Copernican Principle of isotropy and homogeneity of the Universe up to $z\sim0.4$. It should help constrain the contributions to the bulk flows and pull on the Local Group from scales of $>1$ Gpc, if there are any.  It is also very promising in terms of various correlation analyses: it should provide a higher S/N of ISW than possible with 2MPZ; it can also be cross-correlated with the CMB lensing signal to constrain e.g.\ non-Gaussianity (\cite[Giannantonio \& Percival 2014]{GP14}). Some other possible applications include those planned with such major photometric redshift surveys as DES, for instance measuring angular BAOs in redshifts shells to constrain expansion history (see e.g.\ \cite[Blake \& Bridle 2005]{BB05} for an early discussion of such applications). This is certainly not an exhaustive list; as we are planning to publicly release the final photometric redshift catalog, the community will be able to apply many other ideas to these data.

\vspace{1mm}
\textbf{\textit{...and beyond: quasars and deep WISE data.}}
Except for the WISE$\times$SCOS \mbox{$z\sim0.2$} galaxy catalog described here, we are also planning to compile a quasar photo-z sample based on the same data. Our present estimates show that we should be able to identify about 2 million quasars on $>3\pi$ sr (\cite[Krakowski \etal\ 2014, in prep.]{Krakowski14}); it remains to verify how accurate quasar photo-z's will be possible with the limited photometric information we have. In addition, as far as WISE itself is concerned, its all-sky $5\sigma$ limit is at $\sim17$ mag (Vega) in the $W1$ band, which makes it useful to probe additional redshift layers beyond those available through the cross-match with SCOS. WISE itself is also a great AGN/quasar repository as already shown by several authors. All these properties give great promise for future cosmological studies that we are envisaging to undertake.

\vspace{1mm}
\begin{footnotesize}
\textbf{Acknowledgments.} We thank the Wide Field Astronomy Unit at the IfA, Edinburgh for archiving the 2MPZ catalog, which can be accessed at \url{http://surveys.roe.ac.uk/ssa/TWOMPZ}. We made use of data products from the Two Micron All Sky Survey, Wide-field Infrared Survey Explorer, SuperCOSMOS Science Archive, Sloan Digital Sky Survey and Galaxy And Mass Assembly catalogs. Special thanks to Mark Taylor for his wonderful TOPCAT software, \url{http://www.starlink.ac.uk/topcat/}. The financial assistance of the South African National Research Foundation (NRF) towards this research is hereby acknowledged. MB was partially supported by the Polish National Science Center under contract \#UMO-2012/07/D/ST9/02785.
\end{footnotesize}

\end{document}